\documentclass{iaus}
\usepackage{graphicx}

\title[Estimation of distances using tidal streams] 
{Estimation of distances within the MW using tidal streams}

\author[Daniele S. M. Fantin]   
{Daniele S. M. Fantin$^1$}

\affiliation{$^1$Centro de Investigaciones de Astronom\'{i}a, 
    Apdo. Postal 264, M\'{e}rida 5101-A, Venezuela\\
    email: {\tt daniele.fanta21@gmail.com}}

\pubyear{2012}
\volume{289}  
\pagerange{???}
\jname{Advancing the physics of cosmic distances}
\editors{Richard de Grijs \& Giuseppe Bono}

\begin{document}

\maketitle

\begin{abstract}
During the past 20 years, numerous stellar streams have been discovered in both the Milky Way and the Local Group. 
These streams have been tidally torn from orbiting systems, which suggests that most should roughly trace the orbit of their progenitors around the Galaxy. 
As a consequence, they play a fundamental role in understanding the formation and evolution of our Galaxy. 
This project is based on the possibility of applying a technique developed by Binney in 2008 to various tidal streams and overdensities in the Galaxy. 
The aim is to develop an efficient method to constrain the Galactic gravitational potential, to determine its mass distribution, and to test distance measurements. 
Here we apply the technique to the Grillmair \& Dionatos cold stellar stream. 
In the case of noise-free data, the results show that the technique provides excellent discrimination against incorrect potentials and that it is possible to predict the heliocentric distance very accurately. 
This changes dramatically when errors are taken into account, which wash out most of the results. 
Nevertheless, it is still possible to rule out spherical potentials and set constraints on the distance of a given stream.

\keywords{Galaxy: kinematics and dynamics, Galaxy: structure, Galaxy: halo, gravitation, methods: analytical, stars: distances}

\end{abstract}

\firstsection

\section{Introduction}

Establishing a reliable model for the Milky Way (MW) is still one of the main goals of modern astrophysics. 
To accomplish this feat, we need to determine precisely the mass distribution in the MW’s outer regions, which are mainly dominated by dark matter.
Tidal streams, i.e., collections of stars that have been gravitationally stripped from a satellite, usually a dwarf galaxy or globular cluster, are very powerful tools to pursue this goal, because they allow to constrain the Galactic gravitational potential, and thus probe the shape of the dark matter halo (Ibata \textit{et al.} 2001b; Helmi 2004; Fellhauer \textit{et al.} 2006).
\\
In the recent past, several tidal streams and stellar substructures have been discovered in the MW’s halo (for a summary, see Grillmair 2010). 
One of their advantages is that if the measurements of the components’ six phase-space coordinates are reasonably accurate, even a single stream can be sufficient for constraining the potential. 
Unfortunately, in most cases observations provide only three phase-space coordinates (usually the Galactic coordinates and the line-of-sight velocity), with errors which vary significantly from stream to stream. 
The other three coordinates (heliocentric distance and proper motion components) are much harder to obtain, and even when they are measured, their accuracy is usually low.

\section{The Method}

Given the incompleteness of the data, it is difficult to develop efficient orbit-fitting techniques. 
One of the standard approaches is to adopt a gravitational potential and seek an orbit in this potential that is consistent with the data. 
It is based on the assumption that tidal streams are composed of stars that are on closely related orbits, and in particular that they roughly trace the underlying orbit of the progenitor system. 
It reconstructs an orbit through the Galaxy which is consistent with measurements, exploiting the basic principle that if the reconstructed orbit violates the equations of motion, it will also violate energy conservation. 
The technique, developed by Binney (2008), works as follows. 
Given the coordinates of a section of a stream on the sky, $l(u)$ and $b(u)$, the corresponding line-of-sight velocities, $v_{los}(u)$, and a trial potential for the MW, $\phi_{t}(r)$, the missing phase-space coordinates (proper motion components and the heliocentric distance $r$) can be recovered. 
The quantity $u$ is the distance on the sky down the projected orbit, which is integrated in the Miyamoto-Nagai potential (Miyamoto \& Nagai 1975).
Dynamical orbits relevant to the system are identified by computing the variation in the rms energy, $\Delta E \,$, along the track. 
If $\phi_{t} = \phi_{0}$, where $\phi_{0}$ is the true potential of the Galaxy, the recovered phase-space coordinates are consistent with conservation of energy, while if the trial and true potentials differ, energy conservation is violated.

This technique has never been applied to any observational data. 
To test its diagnostic power, we apply it to the Grillmair \& Dionatos cold stellar stream (GD-1; Grillmair 2006; Grillmair \& Dionatos 2006a; Willett \textit{et al.} 2009; Koposov \textit{et al.} 2010). 
This stream is extremely narrow (width less than $0.25^{\circ}$) and it spans $60^{\circ}$ across the sky. 
These dimensions, combined with the fact that no progenitor is present and that it is relatively close to the Sun ($7\,$kpc $\lesssim r_{0} \lesssim 10\,$kpc), makes GD-1 a very good system for orbital modelling.


\section{Results}

\begin{figure*}[t]
\centering
\begin{minipage}{0.495 \textwidth}
\centering
\includegraphics[height=0.25\textheight,width=0.95\textwidth]{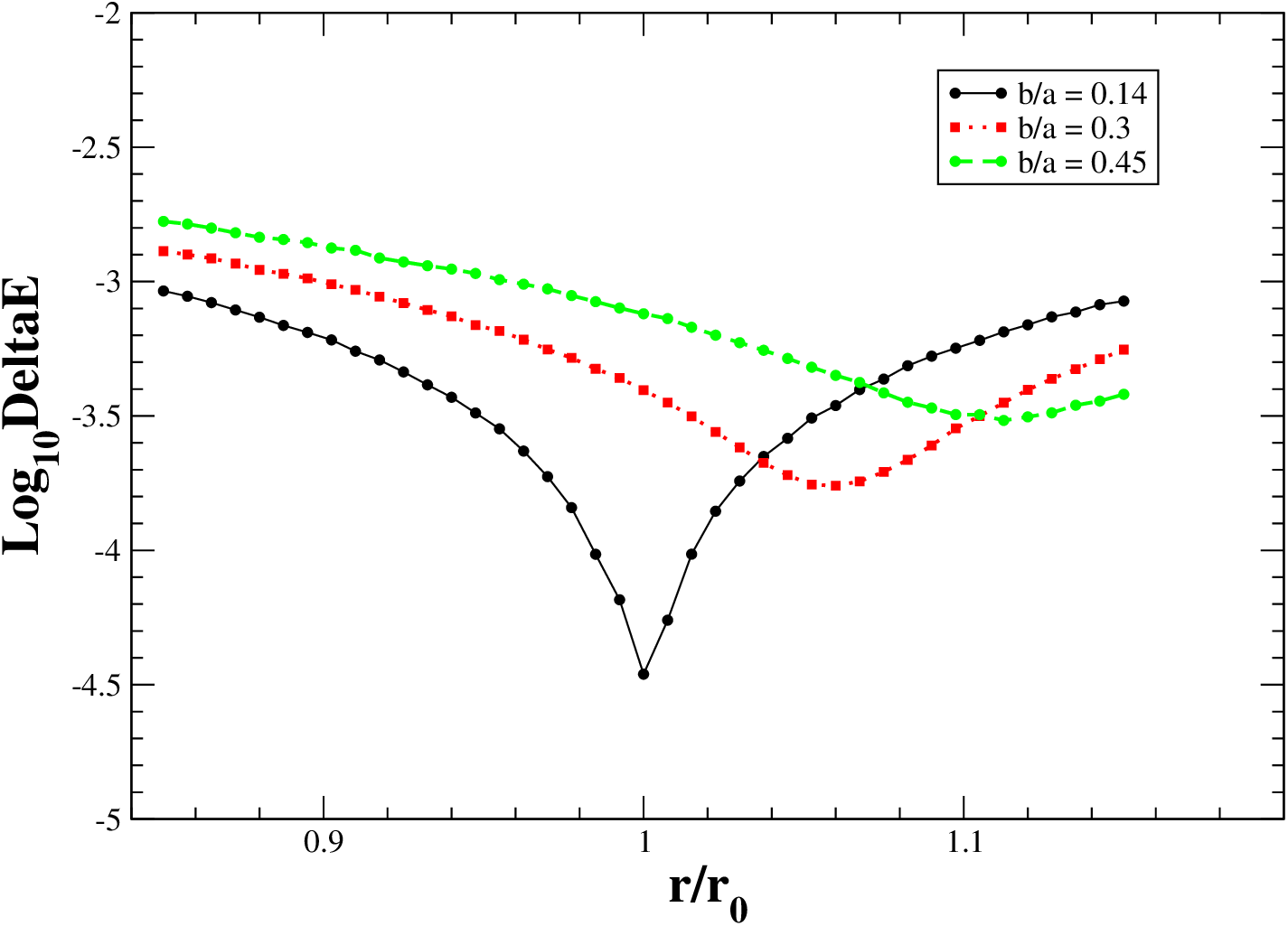}
\vspace{0.6cm}
\end{minipage}
\begin{minipage}{0.495 \textwidth}
\centering
\includegraphics[height=0.25\textheight,width=0.95\textwidth]{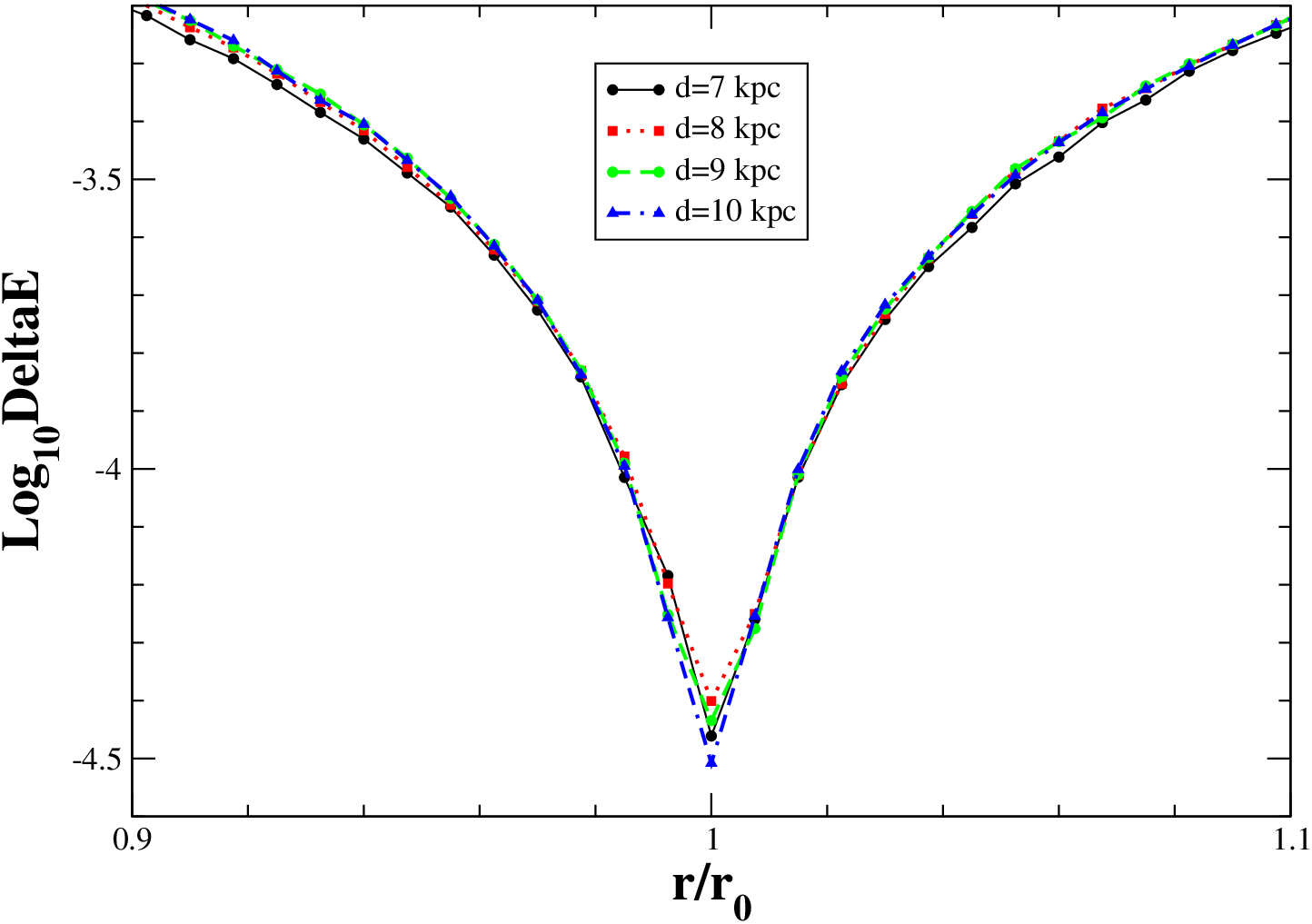}
\vspace{0.6cm}
\end{minipage}
\caption{\textit{(a)} Logarithm of the rms variation in the energy of the reconstructed orbit as a function of the ratio of the initial heliocentric distance, $r$, to its true value, $r_{0}$. 
The value assumed here is $r = 7$ kpc. 
The black, solid curve shows the result obtained when the reconstruction of the orbit employs a flat potential, while the dotted (red) and dashed (green) curves are for rounder ones. 
\textit{(b)} As Fig. \ref{fig1}a, but assuming a flat potential ($b/a = 0.14$) and considering four different test distances: $r = 7$ kpc (black, solid line), $r = 8$ kpc (red, dotted), $r = 9$ kpc (green, dashed), and $r = 10$ kpc (blue, dash-dotted).}
\label{fig1}
\end{figure*}
The method needs the Galactic coordinates and the line-of-sight velocities of two consecutive components of the stream. 
We have applied it to the three available sets of points in Table 1 of Willett \textit{et al.} (2009), and here we show the results obtained for couple 5–6.
Fig. \ref{fig1}a shows the logarithm of the rms variation in the energy when the orbit is reconstructed from an assumed initial distance  $r=7$ kpc, a value within the observational uncertainties. 
Noise-free data are used and different potentials are considered, including $b/a = 0.14,\ 0.3$ and $0.45$.
The figure illustrates the method’s predictive power: when a flat potential is assumed ($b/a = 0.14$; black, solid line), a sharp minimum is recovered. 
This minimum is located exactly at $r = r_{0}$ , suggesting that the energy of the orbit located at $7$ kpc is conserved. 
On the other hand, the orbits reconstructed using more spherical potentials ($b/a = 0.3$, red, dotted line; $b/a = 0.45$, green, dashed one) both show offsets in their $\Delta E \,$ minima, which no longer occur at $r = r_{0}$, and a decrease in their sharpness. 

Fig. \ref{fig1}b shows $(\log \Delta E, r/r_{0})$ for $r=7,\,8,\,9\,$ and $10\,$kpc for a fixed Galactic potential characterized by $b/a = 0.14$. 
The presence of four minima suggests that the energy is conserved along all reconstructed orbits, while their sharpness tells us that they are all realistic approximations of the distance to the stream. 
A more careful analysis shows that the minima produced for $r = 7$ kpc (black, solid curve) and $r = 10$ kpc (blue, dot-dashed) are slightly deeper, and therefore more probable, than those for $r = 8$ kpc (red, dotted) and $r = 9$ kpc (green, dashed). 
In summary, Figs \ref{fig1}a and b show that, when data without errors are considered, the algorithm provides excellent discrimination against incorrect potentials. 
In addition, it can predict the heliocentric distance very accurately once a realistic potential is assumed.

\begin{figure*} [t]
\centering
\begin{minipage}{0.495 \textwidth}
\centering
\includegraphics[height=0.25\textheight,width=0.95\textwidth]{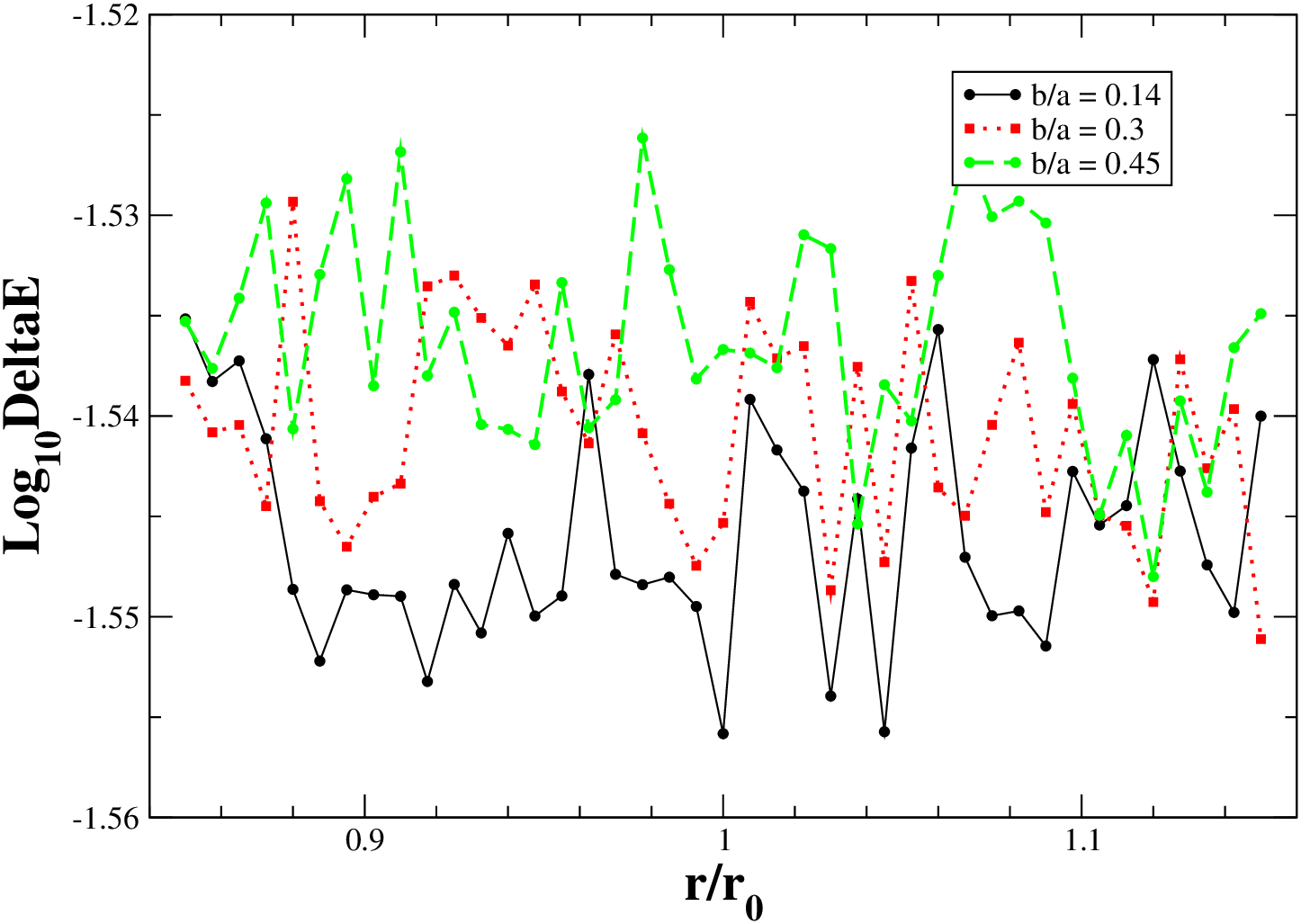}
\end{minipage}
\begin{minipage}{0.495 \textwidth}
\centering
\includegraphics[height=0.25\textheight,width=0.95\textwidth]{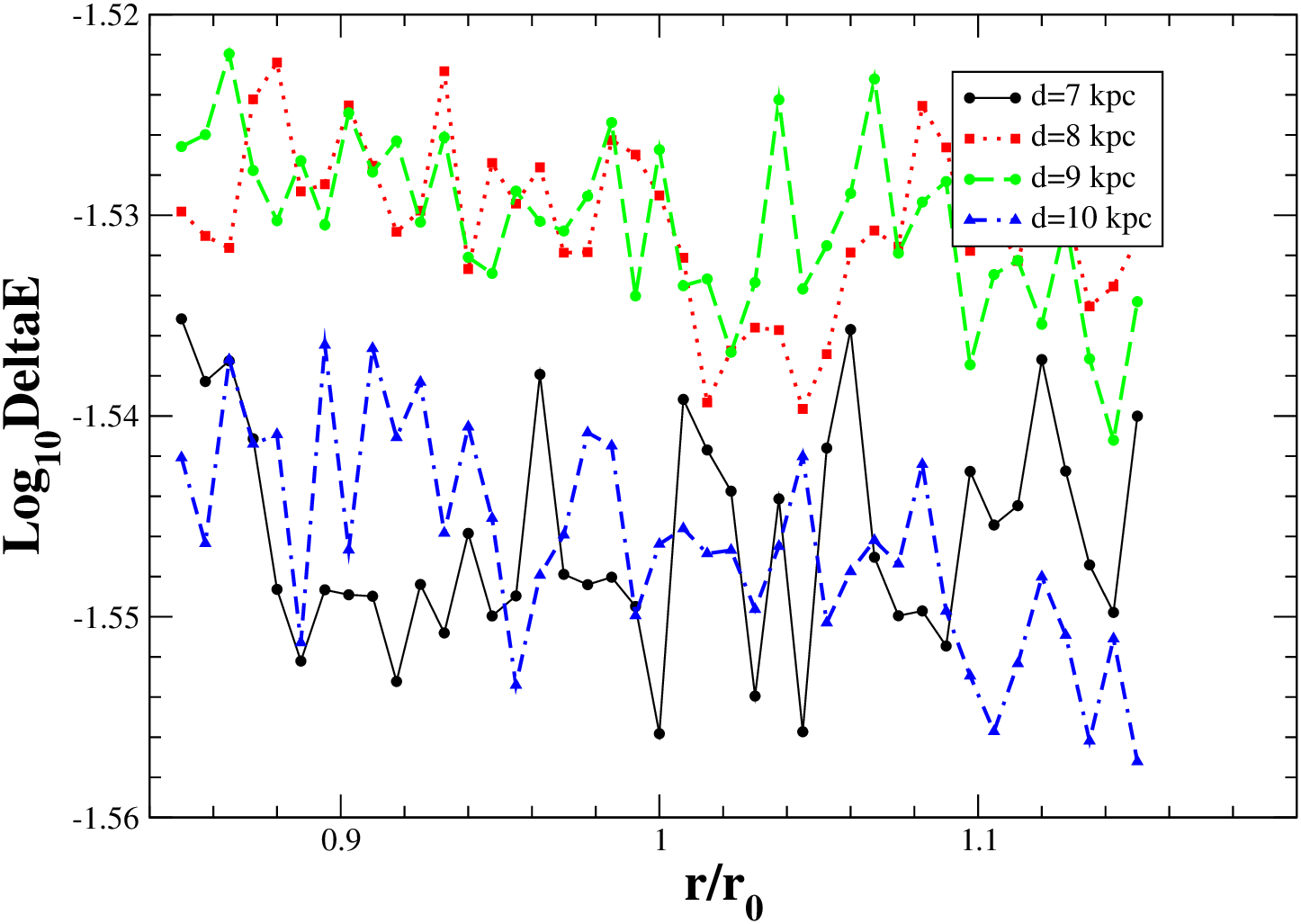}
\end{minipage}
\caption{As Fig. \ref{fig1}, but including the observational errors ($\delta b = \delta l = 0.2^{\circ}$, $\delta v_{los} =2$ km s$^{-1}$).}
\label{fig2}
\end{figure*}

Until now, we assumed that the data were noise-free, but what happens if we reconstruct the orbit of the stream taking into account the observational errors?
Figures \ref{fig2}a and b present the same results as Figs \ref{fig1}a and b, with the only difference that now the input data include errors both in Galactic coordinates ($\delta b = \delta l = 0.2^{\circ}$) and in the line-of-sight velocity ($\delta v_{los} =2$ km s$^{-1}$). 
Fig. \ref{fig2}a shows that the technique, when applied to data, loses most of its predictive power: the results are very noisy and the rms variation in the energy decreases by almost three orders of magnitude compared to Fig. \ref{fig1}a.
The sharp minimum of Fig. 1a is no longer present, and it is not possible to strongly constrain the shape of the potential. 
Nevertheless, if we look in more detail at the solid curve, which corresponds to a $b/a = 0.14$ potential, it still exhibits a small minimum. 
The curve first decreases, then reaches a plateau for $0.9\lesssim r/r_{0} \lesssim 1.05$, and eventually starts to rise at $r/r_{0} \simeq 1.08$. 
Another clue that the flat potential is a reasonable approximation of the real curve is given by the fact that $\log \Delta E$ of the reconstructed orbit reaches its minimum  at $r/r_{0} = 1$, exactly, where we would expect it if the distance $r = 7$ kpc were correct.
On the other hand, the potentials corresponding to $b/a=0.3,\,0.45$ do not show any evidence of a minimum.

Finally, we include the observational errors, fix the Galactic potential ($b/a=0.14$), and calculate the orbit for $7\,$kpc $\leq r \leq 10\,$kpc. 
The results are shown in Fig. \ref{fig2}b. 
The orbits associated with $r = 8$ kpc (red, dotted curve), $r = 9$ kpc (green, dashed), and $r = 10$ kpc (blue, dot-dashed) can be rejected because they do not conserve energy. 
On the other hand, the solid curve exhibits a minimum, as we already saw in Fig. \ref{fig2}a. 
This suggests that the distance $r = 7$ kpc cannot be ruled out, as in the previous cases. 
We conclude that once the observational errors are considered, the quality of the results deteriorates very quickly. 
Nevertheless, the technique still allows us to place constraints on the shape of the potential, which is probably flat, or at least characterized by $b/a < 0.3$, and set constraints on the distance to the stream.

\section{Summary}

We have tested the method developed by Binney (2008) by applying it to the GD-1 stream. 
The aim was to verify how efficiently the method constrains the MW’s gravitational potential and predicts the heliocentric distance to the stream. 
The results show that the method has very good diagnostic power when noise-free data are assumed. 
If observational errors are included, the technique loses most of its power to identify dynamical orbits. 
Nevertheless, it can still constrain the shape of the MW’s potential, yielding $b/a<0.3$ as upper limit and $b/a = 0.14$ as most probable value. 
It also suggests that $r = 7 kpc$ is the most probable heliocentric distance to the GD-1 stream. 
To test how much the results improve once more precise data are assumed, in the near future we plan to apply the technique to the GD-1 data of Koposov \textit{et al.} (2010) and include simulated \textit{Gaia} errors.

\section{Acknowledgments}

DSMF gratefully acknowledges support from doctoral grant of the Academia Nacional de Ciencias F\'{\i}sicas, Matemáticas y Naturales of Venezuela and CIDA.


\begin{thebibliography}{}

\bibitem[Binney (2008)]{Binney08}
{Binney, J.} \  2008, \textit{MNRAS} (Letters), 386, 47 

\bibitem[Fellhauer et al(2006)]{Fellhauer+06}
Fellhauer, M., Belokurov, V., Evans, N. W., et al.\ 2006, \textit{ApJ}, 651, 167 

\bibitem[Grillmair(2010)]{Grillmair10}
Grillmair, C. J.\ 2010, \textit{Galaxies and their Masks}, 247 

\bibitem[Grillmair Dionatos (2006)]{Grillmair+06} 
Grillmair, C. J., \& Dionatos, O.\ 2006, \textit{ApJ}, 643, L17

\bibitem[Grillmair(2006)]{Grillmair06}
Grillmair, C.~J.\ 2006, \textit{ApJ} (Letters), 645, L37 

\bibitem[Helmi(2004)]{Helmi04}
Helmi, A.\ 2004, \textit{ApJ} (Letters), 610, L97 

\bibitem[Ibata et al.(2001)]{Ibata+01}
Ibata, R., Irwin, M., Lewis, G., Ferguson, A. M. N., \& Tanvir, N.\ 2001, \textit{Nature}, 412, 49 

\bibitem[Koposov et al.(2010)]{Koposov+10} 
Koposov, S.E., Rix, H. W., \& Hogg, D.W.\ 2010, \textit{ApJ}, 712, 260 

\bibitem[Miyamoto Nagai(1975)]{Miyamoto+75}
Miyamoto, M., \& Nagai, R.\ 1975, \textit{PASJ}, 27, 533 

\bibitem[Willett et al.(2009)]{Willett+09}
Willett, B.A., Newberg, H.J., Zhang, H., Yanny, B., \& Beers, T.C.\ 2009, \textit{ApJ}, 697, 207 

\end{thebibliography}
\end{document}